\documentclass[a4paper,prb,aps,epsfig,twocolumn,showpacs,preprintnumbers,superscriptaddress,float,amsmath,amssymb]{revtex4-1}
\usepackage{graphicx}
\usepackage{dcolumn}
\usepackage{bm}
\usepackage{booktabs}

\begin{document}

\title{Anomalous Hall effect and current spin polarization in Co$_2$FeX (X = Al, Ga, In, Si, Ge, and Sn) 
Heusler compounds: A systematic {\it ab initio} study}

\author{Hung-Lung Huang}
\affiliation{Department of Physics, National Taiwan University, Taipei 10617, Taiwan}
\author{Jen-Chuan Tung}\email{jenchuan.tung@gmail.com}
\affiliation{Department of Physics, National Taiwan University, Taipei 10617, Taiwan}
\affiliation{Graduate Institute of Applied Physics, National Chengchi University, Taipei 11605, Taiwan}
\affiliation{Center for General Education, China Medical University, Taichung 40402, Taiwan}
\author{Guang-Yu Guo}\email{gyguo@phys.ntu.edu.tw}
\affiliation{Department of Physics, National Taiwan University, Taipei 10617, Taiwan}
\affiliation{Graduate Institute of Applied Physics, National Chengchi University, Taipei 11605, Taiwan}

\date{\today}

\begin{abstract}
Co-based Heusler compounds are ferromagnetic with a high Curie temperature and a large
magnetization density, and thus are promising for spintronic applications.
In this paper, we perform a systematic {\it ab initio} study of two principal spin-related
phenomena, namely, anomalous Hall effect and current spin polarization, in Co$_2$Fe-based Heusler 
compounds Co$_2$FeX (X = Al, Ga, In, Si, Ge, Sn) in the cubic L2$_1$ structure 
within the density functional theory with the generalized gradient approximation (GGA).
The accurate all-electron full-potential linearized augmented plane-wave method is used.
First, we find that the spin-polarization of the longitudinal current ($P^L$) in Co$_2$FeX (X = Al, 
Ga, In, Al$_{0.5}$Si$_{0.5}$ and Sn) is $\sim$100 \% even though that of the electronic states 
at the Fermi level ($P^D$) is not. Further, the other compounds also have a high current
spin polarization with $P^L > 85$ \%. This indicates that all the Co$_2$FeX compounds considered
are promising for spin-transport devices. Interestingly, $P^D$ is negative in Co$_2$FeX (X = Si, Ge and Sn),
differing in sign from the $P^L$ as well as that from the transport experiments. Second, the calculated anomalous 
Hall conductivities (AHCs) are moderate, being within 200 S/cm, and agree well with the available experiments 
on highly L2$_1$ ordered Co$_2$FeSi specimen although they differ significantly from the reported experiments
on other compounds where the B2 antisite disorders were present. Surprisingly, the AHC in Co$_2$FeSi 
decreases and then changes sign when Si is replaced by Ge and finally by Sn.
Third, the calculated total magnetic moments agree well with the corresponding experimental ones in all the
studied compounds except Co$_2$FeSi where a difference of 0.3 $\mu_B$/f.u. exists.
We also perform the GGA plus on-site Coulomb interaction $U$ calculations in the GGA+$U$ scheme.
We find that including the $U$ affects
the calculated total magnetic moment, spin polarization and AHC significantly, and in most cases, unfortunately, 
results in a disagreement with the available experimental results. All these interesting findings
are discussed in terms of the underlying band structures.

\end{abstract}


\maketitle

\section{Introduction}

Most Co-based Heusler compounds in the cubic L2$_1$ structure are ferromagnetic with a high Curie
temperature and a large saturation magnetization.\cite{Brown00} Furthermore, many of them were predicted
to be half-metallic\cite{Galanakis02,Kubler07,Kan07,Tung13} and hence are of particular interest for
spintronics. Therefore, the electronic band structure and magnetic properties of the Co-based Heusler
compounds have been intensively investigated both theoretically and experimentally in recent
years\cite{Brown00,Galanakis02,Kubler07,Kan07,Tung13}. For example, the total magnetic moments of these
materials were found to follow the Slater-Pauling type behavior and the mechanism was explained 
in terms of the calculated electronic structures\cite{Galanakis02}. The Curie temperatures of Co-based 
Heulser compounds were also
determined from {\it ab initio} theoretical calculations and the trends were related to the electronic 
structures\cite{Kubler07}.

Half-metallic ferromagnets are characterized by the coexistence of
metallic behavior for one spin channel and insulating behavior for the other,
and their electronic density of states at the Fermi level is completely spin polarized.
Thus, they could in principle offer a fully spin-polarized current and are useful for spin
electronic devices. The possible half-metallicity of the Co-based Heusler compounds has been
intensively investigated 
experimentally\cite{Gercsi06,Karthik07,Karthik07-1,Hamaya12,Imort12,Oki12,Wu13,USPAT13,Dirk13,Makinistian13}
and theoretically\cite{Galanakis02,Kubler07,Kan07,Karthik07,Makinistian13,Tung13}.
In particular, many point-contact Andreev reflection (PCAR) experiments\cite{Gercsi06,Karthik07,Oki12,Dirk13,Makinistian13}
have been carried out on Co$_2$FeSi and its current spin polarization ($P^L$) was found to vary from
45 \% to 60 \%, depending on the substrate and the quality of the contact.  
A higher $P^L$ of $\sim$80 \% was reported in a nonlocal spin valve (NLSV) experiment\cite{Hamaya12} on Co$_2$FeSi. 
The spin Hall effect experiment\cite{Oki12} showed that the $P^L$ of Co$_2$FeSi is positive.
However, the measured positive current spin polarization is at odds with the predictions
of the negative spin polarization of the electronic states at the Fermi level (static spin polarization)
($P^D$) from the {\it ab initio} calculations\cite{Wur05,Makinistian13} with the local density approximation (LDA)
or generalized gradient approximation (GGA)\cite{Perdew96}.
Furthermore, the calculated spin magnetic moment ($m_s^{tot}$) was found to differ by nearly 10 \% 
from the measured total magnetic moment ($m^{tot}$) of $\sim$6 $\mu_B$/f.u.\cite{Wur05} 
It was argued that the Co-based Heusler compounds are strongly correlated systems and 
hence should be better described by, e.g., the LDA/GGA plus onsite Coulomb Repulsion ($U$) (LDA/GGA+$U$) 
approach\cite{Lie95}. The GGA+$U$ calculations\cite{Wur05} indeed would give rise to a $m_s^{tot}$ of 6 $\mu_B$/f.u.
and a positive $P^D$ of 100 \% (i.e., being half-metallic). However, the calculated $P^D$
is much larger than the measured spin polarization. Nevertheless, as pointed out recently 
in Ref. \onlinecite{Makinistian13}, the spin polarization measured in transport experiments such as
PCAR and NLSV experiments should be compared with the theoretical current spin polarization ($P^L$) rather than
static spin polarization ($P^D$). Therefore, one of the principal purposes of this work is to understand
the measured spin polarization in all the Co$_2$FeX compounds by performing a systematic {\it ab initio}
GGA study of both the current and static spin polarizations as well as the total magnetic moment in these compounds. 
Indeed, we find that the calculated $P^L$ values for the Co$_2$FeX compounds agree with the available
experimental results not only in sign but also in magnitude, while the calculated $P^D$ values
for Co$_2$FeX (X = Si, Ge and Sn) are wrong even in sign. 

The anomalous Hall effect (AHE), discovered in 1881 by Hall\cite{Hal81}, is an archetypal spin-related transport 
phenomenon and hence has recently received renewed attention\cite{Nagaosa10}. 
Indeed, many {\it ab initio} studies on the AHE in elemental ferromagnets\cite{Yao04,Rom09,Fuh11,Tung12} 
and intermetallic alloys\cite{zen06,He12} have recently been reported. However, {\it ab initio} 
investigations into the AHE in the Heusler compounds have been few\cite{Kubler12,Tung13,Tur14}. 
Interestingly, Co$_2$MnX (X = Al, Ga and In) were recently predicted\cite{Tung13} to have a large
intrinsic anomalous Hall conductivity (AHC) in the order of $\sim$1000 S/cm, and thus could find
applications in magnetization sensors\cite{Vidal11}. Therefore, another principal purpose of this work is to understand 
the AHE in all the Co$_2$Fe-based Heusler compounds and the results may help experimental
search for the Heusler compounds with large AHE for applications. Furthermore, by comparison of
the calculated AHC as well as the current spin polarization and total magnetic moment with 
the measured ones, one could have a comprehensive assessment of whether or not the 
Co$_2$FeX compounds are strongly correlated systems that would require the GGA+$U$ approach.

In nonmagnetic materials where the numbers of the spin-up and spin-down electrons
are equal, the opposite transverse currents caused by the applied electric field 
would result in a pure spin current, and this is known as the intrinsic spin
Hall effect (SHE).\cite{Mur03} The pure spin current is dissipationless\cite{Mur03} and is
thus important for the development of low energy-consumption nanoscale spintronic devices\cite{Liu12}.
We note that high spin-polarization ($P^L$) of the charge current ($I_C$) from the electrode is essential
for large giant magnetoresistance (GMR)\cite{Gru86,Bai88} and tunneling magnetoresistance (TMR)\cite{Moo95,Parkin04}. 
However, since the current-induced magnetization switching results from the transfer
of spin angular momentum from the current carriers to the magnet\cite{Slo96}, large spin current ($I_S$)
would be needed for the operation of the spin-torque switching-based nanodevices\cite{Slo96,Mye99}, i.e.,
a large ratio of spin current to charge current [$\eta = |(2e/\hbar) I_S/I_C|$], would be crucial. For
ordinary charge currents, this ratio $\eta$ varies from 0.0 (spin unpolarized current) to 1.0 (fully spin
polarized current). Interestingly, $\eta$ can be larger than 1.0 for the Hall currents and is
$\infty$ for pure spin current. Fascinatingly, spin-torque switching of ferromagnets driven by pure
spin current from large SHE in tantalum has been recently reported\cite{Liu12}. 
Therefore, it might be advantageous to use the Hall current from ferromagnets for magnetoelectronic devices, rather than the
longitudinal current. Another purpose of this work is therefore to investigate the nature and
spin-polarization of the Hall current in the Co$_2$Fe-based Heusler compounds for 
possible spintronic applications.

The rest of this paper is organized as follows. In the next section, we briefly describe the theories of 
the intrinsic anomalous and spin Hall conductivities as well as Hall and longitudinal current spin polarizations. 
We will also introduce the full-potential relativistic band theoretical method used and give the computional
details. In Sec. III, the calculated magnetic moments, intrinsic Hall conductivities and current spin polarizations
of all the studied Co$_2$FeX compounds will be reported in subsections III (A), (B) and (C), 
respectively. The theoretical results will be 
analyzed in terms of the underlying band structures and also compared with the available experimental ones.   
In subsection III (D), the results from the GGA+$U$ calculations will be presented to 
examine the effect of including the semi-empirical on-site Coulomb interaction on the calculated
physical properties of the Co$_2$FeX compounds. Finally, conclusions drawn from this work will be
given in Sec. IV.
 
\section{Theory and computational method}

We first perform the self-consistent electronic structure calculations for the Co$_2$FeX compounds 
within the density functional theory with the GGA for 
the exchange correlation potential\cite{Perdew96}. Since all the intrinsic Hall effects are caused by 
the relativistic electron spin-orbit interaction, the spin-orbit coupling (SOC) is included in the 
present {\it ab initio} calculations.  We use the highly accurate full-potential linearized 
augmented plane wave (FLAPW) method, as implemented in the WIEN2K code\cite{wien2k02}. 
The wave function, charge density, and potential were expanded in terms of the spherical harmonics inside
the muffin-tin spheres and the cutoff angular moment ($L_{max}$) used is 10, 6 and 6, respectively. 
The wave function outside the muffin-tin spheres is expanded in terms of the augmented plane waves (APWs) 
and a large number of APWs (about 70 APWs per atom, i. e., the
maximum size of the crystal momentum $K_{max}=8/R_{mt}$) were included in the present calculations. 
The improved tetrahedron method is used for the Brillouin-zone integration\cite{Blochl94}. 
To obtain accurate ground state charge density as well as spin and orbital magnetic moments, 
a fine 27$\times$27$\times$27 grid with 1470 $k$-points in the irreducible Brillouin zone wedge (IBZW) is used.
The self-consistent cycles were terminated when the integrated charge density variation became
less than $10^{-5}$ e.

We consider the Co$_2$FeX Heusler compounds in the fully ordered cubic L2$_1$ structure. The available
experimental lattice constants\cite{2007_APL90-172501} are used for all the considered Co$_2$FeX (X = Al, Ga, In, Si, Ge, Sn)
Heusler alloys except Co$_2$FeAl$_{0.5}$Si$_{0.5}$, Co$_2$FeIn and Co$_2$FeSn, as listed in Table I.
Since the experimental lattice constant for Co$_2$FeIn is not available, we determine the lattice 
constant for Co$_2$FeIn theoretically, also by using the FLAPW method, as described in the preceding paragraph. 
We also study the L2$_1$ Co$_2$FeAl$_{0.5}$Si$_{0.5}$ alloy and model it by
the virtual crystal approximation (VCA), i.e., the Al/Si site is occupied by a virtual atom 
with the atom number $Z = 0.5Z_{Al} + 0.5Z_{Si}$, where $Z_{Al}$ and $Z_{Si}$ are the Al and Si 
atomic numbers, respectively. The lattice constant of 5.689 \AA, which 
is the average of the experimental lattice constants of Co$_2$FeAl (5.737 \AA) and Co$_2$FeSi (5.640 \AA),
is used for Co$_2$FeAl$_{0.5}$Si$_{0.5}$ because the lattice constant of the Co$_2$FeAl$_{x}$Si$_{1-x}$ alloy
was reported to depend linearly on the Al concentration ($x$)\cite{Nakatani07}. 
Note that in fact we have determined the lattice constants theoretically also for the 
Co$_2$FeX (X = Al, Ga, Si, Ge) compounds. The theoretical lattice constants for these 
compounds differ from the experimental values by less than 1 \%. As a result, the physical properties 
of these compounds calculated using the experimental and theoretical 
lattice constants differ only slightly. Therefore, for simplicity,
we present only the physical properties of these compounds calculated using the experimental
lattice constants in the next section. However, the theoretical lattice constant (6.013 \AA) 
of Co$_2$FeSn is 2.4 \% larger than the experimental one (5.87 \AA)\cite{tanaka12} perhaps
because the prepared Co$_2$FeSn films contained only a low degree of L2$_1$ order. 
Therefore, the theoretical lattice constant is used for Co$_2$FeSn (Table I).

\subsection{Anomalous and spin Hall conductivities}
The intrinsic anomalous and spin Hall conductivities of a solid can be evaluated by using 
the Kubo formalism.\cite{Guo95,Yao04,Guo05} 
Here we first calculate the imaginary part of the off-diagonal elements of the optical conductivity. 
Then we obtain the real part of the off-diagonal elements from the imaginary part 
by a Kramers-Kronig transformation. The intrinsic AHC ($\sigma^{A}_{xy}$) is the 
static limit of the off-diagonal element of the optical conductivity $\sigma^{(1)}_{xy}(\omega=0)$.
If we now replace the charge current operator $-e\hat{\bf v}$ with the spin current operator 
$(\hbar/4)\{\Sigma_z,\hat{\bf v}\}$ and repeat the calculation\cite{Guo95}, we will obtain the intrinsic
spin Hall conductivity (SHC) ($\sigma^{S}_{xy}$). We note in passing that alternatively, one could 
also calculate $\sigma^{A}_{xy}$ ($\sigma^{S}_{xy}$) by an integration of the (spin) Berry curvature
over the Brillouin zone\cite{Yao04,Guo08,Guo14}. Nevertheless, the two methods were found to be 
numerically equivalent\cite{Yao04,Guo08,Guo14}. 

A dense $k$ point mesh would be needed for obtaining accurate AHC and SHC\cite{Yao04,Guo05}. 
Therefore, we use several fine $k$-point meshes with the finest $k$-point mesh being 58$\times$58$\times$58 which has
8125 $k$-points in the IBZW. We calculate the AHC and SHC 
as a function of the number ($N_k$) of $k$-points in the first Brillouin zone. The calculated AHC 
($\sigma^{A}_{xy}$) and SHC ($\sigma^{S}_{xy}$) versus the inverse of the $N_k$ 
are then plotted and fitted to a polynomial to get the converged theoretical $\sigma^{A}_{xy}$ 
and $\sigma^{S}_{xy}$ (i.e., the extrapolated value at $N_k = \infty$) (see Refs. \onlinecite{Fuh11} 
and \onlinecite{Tung12}). Furthermore, to ensure that the $\sigma^{(1)}_{xy}(\omega=0)$ via
the Kramers-Kronig transformation is accurate, the energy bands
up to 5.5 Ry are included in the calculation of $\sigma^{(2)}_{xy}(\omega)$. 

\subsection{Current spin polarization}
The spin polarization of a magnetic material is usually described in terms of the spin-decomposed 
densities of states (DOSs) at the Fermi level ($E_F$) as follows
\begin{equation}
P^{D} =\frac{N_{\uparrow}(E_F)-N_{\downarrow}(E_F)}{N_{\uparrow}(E_F)+N_{\downarrow}(E_F)},
\end{equation}
where $N_{\uparrow}(E_F)$ and $N_{\downarrow}(E_F)$ are the spin-up and spin-down DOSs at the $E_F$, respectively. 
This static spin polarization $P^{D}$ would then vary from -1.0 to 1.0 only. For the half-metallic materials, $P^{D}$ equals 
to either -1.0 or 1.0. As mentioned above, the spin polarization $P^D$ 
defined by Eq. (1) is not necessarily the spin polarization of the transport currents measured in experiments. 
Indeed, the spin-polarizations measured by using different experimental techniques could differ 
significantly\cite{Sou98,Ter99,Dow11,Koz14}.
From the viewpoint of spintronic applications, only the current spin polarization instead of the $P^D$, counts.

Therefore, in this work, we further calculate the spin polarization of both the longitudinal and Hall currents, as described
below. Here, we calculate the longitudinal electric conductivities ($\sigma_{\uparrow}, \sigma_{\downarrow}$) 
for spin-up and spin-down electrons divided by the corresponding Drude relaxation times ($\tau_{\uparrow}, \tau_{\downarrow}$) 
(i.e., $\sigma_{\uparrow}/\tau_{\uparrow}, \sigma_{\downarrow}/\tau_{\downarrow}$) within 
the semi-classical Boltzmann transport theory, as implemented in BoltzTrap code\cite{Boltz06}. 
In the present calculations, the relaxation time is assumed to be independent of energy, $k$-point and spin direction
(i.e., $\tau_{\uparrow} = \tau_{\downarrow} = \tau$).
Consequently, we can obtain the longitudinal current spin polarization $P^L$ from
\begin{equation}
P^L =\frac {\sigma^{\uparrow}-\sigma^{\downarrow}} {\sigma^{\uparrow}+\sigma^{\downarrow}}\simeq \frac {\sigma^{\uparrow}/\tau-\sigma^{\downarrow}/\tau} {\sigma^{\uparrow}/\tau+\sigma^{\downarrow}/\tau}.
\end{equation}
The underlying scalar-relativistic band structures are calculated by using a fine 36$\times$36$\times$36 mesh 
with 3349 $k$-points in the IBZW.

The spin polarization $P^H$ of the Hall current may be written as\cite{Tung12,Tung13}
\begin{equation}
P^H =\frac{\sigma^{H\uparrow}_{xy}-\sigma^{H\downarrow}_{xy}}{\sigma^{H\uparrow}_{xy}+\sigma^{H\downarrow}_{xy} }
\end{equation}
where $\sigma^{H\uparrow}_{xy}$ and $\sigma^{H\downarrow}_{xy} $ 
are the spin-up and spin-down Hall conductivities, respectively. 
The $\sigma^{H\uparrow}_{xy}$ and $\sigma^{H\downarrow}_{xy}$ can be 
obtained from the calculated AHC and SHC via the relations\cite{Guo14}
\begin{equation}
\sigma^{A}_{xy} = \sigma^{H\uparrow}_{xy}+\sigma^{H\downarrow}_{xy}
\end{equation}
\begin{equation}
-2\frac{e}{\hbar}\sigma^{S}_{xy} =
\sigma^{H\uparrow}_{xy}-\sigma^{H\downarrow}_{xy}.
\end{equation}
Note that, the absolute value of $P^H$ can be greater than 1.0 because the spin-decomposed Hall currents can go either right
(positive) or left (negative). In the nonmagnetic materials, the charge Hall current is zero, and hence, 
$\sigma^{H\uparrow}_{xy}$ $=$ $-\sigma^{H\downarrow}_{xy} $ results in $P^H$$=\infty$. 
Clearly, in the case of Hall currents, the ratio of
the spin current to charge current $\eta = |(2e/\hbar) \sigma^{S}_{xy} / \sigma^{A}_{xy} | = |P^H|$. 

\begin{table*}
\caption{Calculated total spin magnetic moment ($m_s^{tot}$) ($\mu_B$/f.u.), atomic spin ($m_s$) and orbital ($m_o$)
magnetic moments ($\mu_B$/atom) as well as spin-decomposed density of states at the Fermi level [$N^{\uparrow}(E_F)$,
$N^{\downarrow}(E_F)$] (states/eV/f.u.) of all the considered Co$_2$FeX Heusler compounds together with the lattice 
constants $a$ (\AA) used. The available experimental magnetic moments\cite{1983_JMMM38-1,2005_PRB72-184434,2001_Ziebeck,Kumar09} 
(Exp.) are also listed for comparison
with the calculated total magnetic moments ($m^{tot}$) ($\mu_B$/f.u.). In Co$_2$FeAl$_{0.5}$Si$_{0.5}$, 
listed in the bracket is the spin magnetic moment of Si. The orbital magnetic moments for the non-transition metal atoms 
($m_{o}^{X}$) are less than 0.0001 $\mu_B$/atom and hence not listed here. The theoretical total magnetic moment \
($m^{tot}$) is given by $m^{tot} = m_{s}^{tot} + 2m_{o}^{Co} + m_{o}^{Fe}$, which should be compared
with the experimental magnetic moment.}
\begin{center}
\begin{tabular}{ccccc@{\hspace{0.3cm}}c@{\hspace{0.3cm}}c@{\hspace{0.3cm}}c@{\hspace{0.3cm}}c@{\hspace{0.3cm}}ccc}\hline\hline
Co$_2$FeX & $a$ &  & $m^{tot}$ & $m_{s}^{tot}$ & $m_{s}^{Co}$ & $m_{s}^{Fe}$ &
$m_{s}^X$ &$m_{o}^{Co}$ &$m_{o}^{Fe}$ &$N^{\uparrow}(E_F)$&
$N^{\downarrow}(E_F)$\\ \hline
Co$_2$FeAl &5.737\footnotemark[1]&GGA    &5.123 &4.993 &1.229 &2.788 &-0.064 &0.041 &0.048 &0.862 &0.059 \\
           &               &Exp.   &4.96 & & & & & & & & \\
Co$_2$FeGa &5.751\footnotemark[1]&GGA    &5.149 &5.016 &1.206 &2.811 &-0.047 &0.041 &0.051 &0.885 &0.189 \\
           &               &Exp.   &5.13 & & & & & & & & \\
Co$_2$FeIn &5.990          &GGA      & 5.308 & 5.143 & 1.250& 2.885& -0.046& 0.052  & 0.061  & 0.859  & 0.575 \\
Co$_2$FeAl$_{0.5}$Si$_{0.5}$ & 5.689 &GGA  & 5.523 & 5.376 & 1.338 & 2.683 & -0.037 & 0.052 & 0.043 & 0.755 & 0.399 \\
Co$_2$FeSi &5.640\footnotemark[1]&GGA    &5.688 &5.541 &1.388 &2.848 &-0.002 &0.040 &0.067 &0.714 &2.476 \\
           &               &Exp.   &5.97 & & & & & & & & \\
Co$_2$FeGe &5.743\footnotemark[1]&GGA    &5.854 &5.693 &1.422 &2.917 & 0.012 &0.046 &0.069 &0.785 &2.288 \\
           &               &Exp.   &5.90, 5.74 & & & & & & & & \\
Co$_2$FeSn &6.013          &GGA    &5.994 &5.797 & 1.445& 3.021& -0.005& 0.060& 0.079& 0.712 & 2.457 \\
           \hline\hline
\end{tabular}
\end{center}
\footnotetext[1]{Experimental lattice constants\cite{2007_APL90-172501}.}\
\end{table*}

\section{Results and discussion}
\subsection{Magnetic moments and band structure}
Let us first examine the calculated magnetic properties and band structures near the Fermi level of the considered 
Co$_2$FeX Heusler alloys. Since the electronic structure and magnetism in the full Heusler compounds have been 
extensively studied (see, e.g., Refs. \onlinecite{Galanakis02,Kubler07,Kan07} and references therein), 
here we focus on only the salient features which may be related to the anomalous and spin Hall effects as well as 
current spin polarizations to be presented in the next subsections. The calculated total magnetic moment,
total spin magnetic moment, local spin and orbital magnetic moments as well as spin-decomposed DOSs 
at $E_F$ of all the considered Co$_2$FeX Heusler alloys are listed in Table I, 
together with the available experimental total magnetic moments for comparison. 
The total and site decomposed DOSs of three selected Heusler compounds Co$_2$FeAl,
Co$_2$FeAl$_{0.5}$Si$_{0.5}$ and Co$_2$FeSi are displayed in Fig. 1. The scalar relativistic band structures 
of Co$_2$FeAl and Co$_2$FeSi are shown in Figs. 2(a) and 2(c), respectively. 

\begin{figure}
\includegraphics[width=7cm]{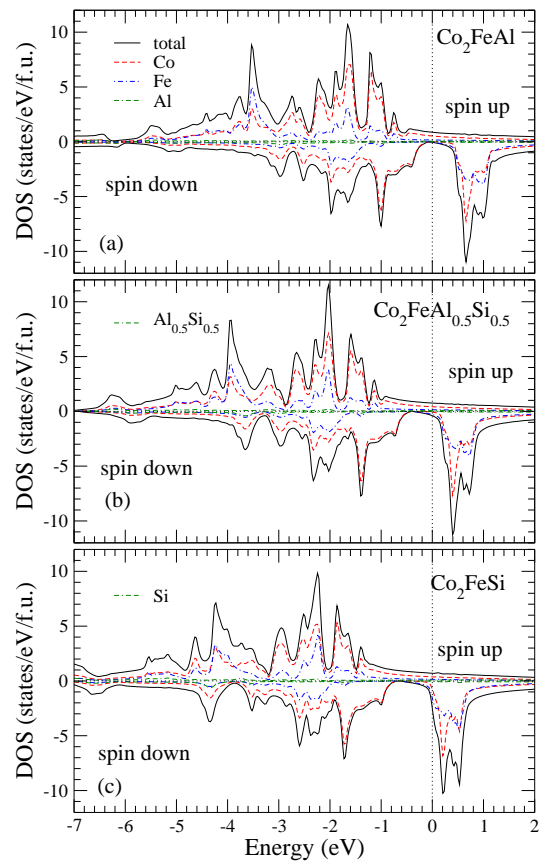}\
\caption{(color online) Total and site decomposed density of states (DOS) for (a) Co$_2$FeAl, (b) Co$_2$FeAl$_0.5$Si$_0.5$ and
(c) Co$_2$FeSi. The Fermi level is at zero.}
\label{dos1}
\end{figure}

Interestingly, the studied Heusler alloys could be separated into two groups according to the calculated DOS 
at $E_F$. In one group, including Co$_2$FeAl, Co$_2$FeGa, Co$_2$FeAl$_{0.5}$Si$_{0.5}$ and Co$_2$FeIn, 
the majority spin state dominates. In the other group, including Co$_2$FeSi, Co$_2$FeGe and Co$_2$FeSn, 
the minority spin state dominates (see Table I). Therefore, the calculated spin polarization ($P^D$) for the 
first group is positive while that of the second group is negative (see Table II). 
It is clear from Fig. 1(a) that there is a band gap near the $E_F$ for the minority spin channel in Co$_2$FeAl
and adding one valance electron can be approximatively treated as raising the $E_F$ by $\sim$0.4 eV. 
This $E_F$ shift is nearly equal to that from Co$_2$MnAl to Co$_2$MnSi in Ref. \onlinecite{Tung13} where 
the calculated $P^D$ for both compounds, however, is positive. This is because the minority gap here 
is small, being $\sim$ 0.2 eV, while that in Co$_2$MnAl is much larger, being $\sim$ 0.8 eV.

Figure 1 and Table I show that from the view point of the calculated band structures, all the considered Heusler compounds 
are not half-metallic, although Co$_2$FeAl is nearly a half-metal because its $E_F$ just touches the bottom of the minority 
spin conduction band [Fig. 2(a)]. Previous GGA calculations\cite{Gerhard07} also predicted that 
of Co$_2$FeAl, Co$_2$FeAl$_{0.5}$Si$_{0.5}$ and Co$_2$FeSi are not half-metallic. 
The DOS spectra for Co$_2$FeAl, Co$_2$FeAl$_{0.5}$Si$_{0.5}$, and Co$_2$FeSi are similar 
and differ only in the location of $E_F$ (see Fig. 1). The DOS spectra of Co$_2$FeGa (Co$_2$FeIn) and
Co$_2$FeGe (Co$_2$FeSn) (not shown here) also look similar, except the location of $E_F$.

Table I indicates that among the considered Heusler compounds, only the total spin magnetic moment $m_s^{tot}$ 
of Co$_2$FeAl and Co$_2$FeGa almost satisfies the so-called generalized Slater-Pauling rule $m_s^{tot} = n_v - 24$ where
$n_v$ is the number of valence electrons\cite{Galanakis02}. This may be expected because none of these compounds
is predicted to be half-metallic here and only Co$_2$FeAl is nearly half-metallic.
Table I also suggests that the calculated $m^{tot}$ agrees well with the available measured one for all the considered
compounds except Co$_2$FeSi. The discrepancy between the calculated and experimental $m^{tot}$ 
is $\sim$0.3 $\mu_B$/f.u. for Co$_2$FeSi but is about 0.1 $\mu_B$/f.u.
or less for all the other compounds (Table I). 

\subsection{Anomalous and spin Hall conductivities}

\begin{table*}
\caption{Calculated anomalous [$\sigma_{xy}^A$ (S/cm)] and spin [$\sigma_{xy}^S$ ($\hbar$S/e cm)] Hall conductivities, 
spin-decomposed Hall conductivities ($\sigma^{H\uparrow}_{xy}$, $\sigma^{H\downarrow}_{xy}$) (S/cm), 
Hall ($P^H$) and longitudinal ($P^L$) current spin polarizations ($\%$) as well as spin polarization of 
the electronic states at the Fermi level $P^{D}$ ($\%$) of all the considered Heusler compounds Co$_2$FeX.  
The available experimental spin polarization and scattering-independent part ($b$)\cite{note2} of the $\sigma_{xy}^A$
are also listed for comparison. Note that $b$ contains both the intrinsic contribution ($\sigma_{xy}^A$) calculated here
and also the extrinsic side-jump contribution ($\sigma_{xy}^{A-sj}$).
}
\begin{center}
\begin{tabular}{ccc@{\hspace{0.5cm}}c@{\hspace{0.5cm}}c@{\hspace
{0.5cm}}c@{\hspace{0.5cm}}c@{\hspace{0.5cm}}c@{\hspace{0.5cm}}c@{\hspace{0.5cm}}}\hline\hline
Co$_2$FeX & &$\sigma^{A}_{xy}$,$b$ &$\sigma^{S}_{xy}$
&$\sigma^{H\uparrow}_{xy}$ &$\sigma^{H\downarrow}_{xy}$ &$P^H$ &$P^{L}$ & $P^D$ \\ \hline
Co$_2$FeAl &GGA     & 39 & 35 & -16 & 55 & -180  & 100 & 87 \\
           &Exp.    & 320$\sim$360\footnotemark[1]   &    &      &      &       &  56\footnotemark[2]  &   \\
Co$_2$FeGa &GGA     &181 & 56 & 35 &147 &  -62  & 98  & 65 \\
           &Exp.    &    &    &      &      &       & 57\footnotemark[3]  &  \\
Co$_2$FeIn &GGA     & 102 &  56  &  -5 & 107    &   -110   &   92 & 20  \\
Co$_2$FeAl$_{0.5}$Si$_{0.5}$    &GGA     &  124  &   74    & -12& 136 & -119 & 92 & 31 \\
           &Exp.    &  -100$\sim$50\footnotemark[7], 352\footnotemark[8]  &    &      &      &       & 60\footnotemark[5]   &  \\
Co$_2$FeSi &GGA     &189 & 24 & 71 &119 & -25  & 86 & -55 \\
           &Exp.    &163\footnotemark[4],300$\sim$600\footnotemark[1] &    &   &   & & 45$\sim$60\footnotemark[5], 80\footnotemark[6] &  \\
Co$_2$FeGe &GGA     &119 &-29 & 89 & 31 &  49  & 89 & -49 \\
           &Exp.    &    &    &    &      &       &      59\footnotemark[3] &  \\
Co$_2$FeSn &GGA     &-78 &-24 &-15 & -63&  -62 & 93 & -55 \\
\hline\hline
\end{tabular}
\end{center}
\footnotetext[1] {Experimental $b$ values from sputtered films with the B2 structure\cite{Imort12}.}\
\footnotetext[2] {Point-contact Andreev reflection experiments\cite{Karthik07-1}.}\
\footnotetext[3] {Point-contact Andreev reflection experiments on Co$_2$FeGa$_x$Ge$_{1-x}$ in the L2$_1$/B2 mixed structure\cite{USPAT13}.}\
\footnotetext[4] {Experimental $b$ value from Co$_2$FeSi single crystals with the L2$_1$ structure\cite{Dirk13}.}\
\footnotetext[5] {Point-contact Andreev reflection experiments\cite{Makinistian13,Gercsi06,Karthik07,Nakatani07}.}\
\footnotetext[6] {Nonlocal spin-valve experiment\cite{Hamaya12}.}\
\footnotetext[7] {Experimental $b$ values from sputtered Co$_2$FeAl$_{0.4}$Si$_{0.6}$ films with the B2 structure\cite{Vil11}.}\
\footnotetext[8] {Experimental $b$ value from sputterd ultrathin Co$_2$FeAl$_{0.5}$Si$_{0.5}$ film with the B2 structure\cite{Wu13}.}\
\end{table*}

The calculated anomalous Hall conductivity $\sigma^{A}_{xy}$ and spin Hall conductivity $\sigma^{S}_{xy}$ 
for all the studied compounds are listed in Table II.
We notice that compared with the AHC of Fe metal\cite{Yao04} and also Co$_{2}$MnX (X = Al, Ga and In)\cite{Tung13}, 
the $\sigma^{A}_{xy}$ of the present Heusler compounds is moderate in magnitude, being within 200 S/cm (Table II).
In fact, the AHC of Co$_{2}$FeX (X = Al, Ga and In) (Table II) is about one order of magnitude smaller than
the corresponding Co$_{2}$MnX (X = Al, Ga and In) (see Table II in Ref. \onlinecite{Tung13}).  
This can be explained in terms of the calculated band structure and also $\sigma^{A}_{xy}$ as a function of
$E_F$ in Co$_{2}$FeAl (Figs. 2a and 2b). Figures 2a and 2c show 
that the spin-up bands near $E_F$ are the highly dispersive
Co $spd$, Fe $spd$ and Al $sp$ hybridized bands while $E_F$ nearly falls within the spin-down band gap. Consequently, 
the  $\sigma^{A}_{xy}$ is rather small (being $\sim$35 S/cm) (see Fig. 2b). However, when $E_F$ is
lowered to below -0.8 eV, $\sigma^{A}_{xy}$ increases dramatically to the values of $\sim$1000 S/cm (Fig. 2b).
These large $\sigma^{A}_{xy}$ values come mainly from the spin-up Co $d$ dominant bands 
in this energy range (Figs. 2a and 2c).
In Co$_{2}$MnAl, the corresponding spin-up Co $d$ dominant bands are higher in energy, and
the $E_F$ is lower because Co$_{2}$MnAl has one fewer valence electron than Co$_{2}$FeAl.
As a result, the $E_F$ sits on the Co $d$ dominant $\sigma^{A}_{xy}$ peak in Co$_{2}$MnAl and thus Co$_{2}$MnAl
has a much larger $\sigma^{A}_{xy}$ (being $\sim$ 1300 S/cm).\cite{Tung13}
This interesting finding suggests a way to chemical composition tuning of the AHC 
in Co$_{2}$Mn$_{1-x}$Fe$_{x}$X (X = Al, Ga and In) alloys.

Interestingly, for the Co$_{2}$FeX (X = Si, Ge and Sn) compounds, the AHC gets reduced when Si is replaced by
Ge and changes sign when Ge is further substituted by Sn. 
Nevertheless, the calculated band structures for the Co$_{2}$FeX (X = Si, Ge and Sn) compounds
look very similar, especilly in the vinicity of $E_F$. Thus, there is no obvious explanation for this
interesting evolution. 
Table II indicates that the $\sigma_{xy}^A$ of Co$_2$FeSi is about five times larger than that of Co$_2$FeAl.
This could be attributed to the band filling effect. Figure 2 shows that in Co$_2$FeSi, due to the additional 
one valence electron, the $E_F$ is raised to the bottom of the Co/Fe $d (e_g)$ dominant bands where $\sigma_{xy}^A$ 
is large (Fig. 2b), thus resulting in a much larger $\sigma_{xy}^A$.

Several AHE experiments on the Co$_{2}$FeX compounds and their alloys
have been carried out.\cite{Dirk13,Imort12,Vil11,Wu13} The derived AHC 
values ($b$)\cite{note2} for Co$_2$FeAl, Co$_{2}$FeAl$_{0.5}$Si$_{0.5}$ and Co$_2$FeSi are listed in Table II.
However, quantitative comparison of the present theoretical calculations with the experimental results
is difficult, because all the samples used in the experiments except Co$_2$FeSi are in the B2 structure
with antisite disorders (Table II). The deduced AHC values depend strongly on the substrates used
and annealing temperatures which control the degree of the B2 antisite disorders and also the defect
concentrations.\cite{Imort12,Vil11} Nevertheless, Table II shows that the calculated $\sigma_{xy}^A$ 
of Co$_{2}$FeSi is in good agreement with the experimental result from the single crystal sample\cite{Dirk13},
indicating the intrinsic AHC $\sigma_{xy}^A$ dominates in Co$_{2}$FeSi single crystals with the L2$_1$
structure. In contrast, the theoretical $\sigma_{xy}^A$ of Co$_{2}$FeAl is one order of magnitude smaller than
the $b$ derived from the experiment\cite{Imort12}. For Co$_{2}$FeAl$_{0.5}$Si$_{0.5}$, the $b$ values
from two different experiments\cite{Vil11,Wu13} are very different, suggesting the important influences 
of the B2 antisite disorder and also the substrate.
We could attribute the pronounced discrepancies between the theoretical (intrinsic) ($\sigma_{xy}^A$)
and experimental $b$ values\cite{Imort12} of Co$_{2}$FeAl and Co$_{2}$FeSi to a significant contribution 
from the impurity side-jump scattering as well as the structural difference.
However, recent {\it ab initio} calculations\cite{Tur14} for Co$_2$CrAl and Co$_2$MnAl indicated that
the B2 antisite disorders tend to significantly reduce the intrinsic AHC $\sigma_{xy}^A$.
Therefore, in the experiments\cite{Imort12} on Co$_2$FeAl and Co$_2$FeSi, 
side-jump mechanism could dominate and thus result in a much larger $b$ than $\sigma_{xy}^A$.

Table II indicates that the calculated $\sigma^{S}_{xy}$ in  Co$_{2}$FeIn and Co$_{2}$FeAl$_{0.5}$Si$_{0.5}$
is about half of the $\sigma^{A}_{xy}$ and their Hall current spin polarization ($P^H$) 
is nearly 100 \%. In a half-metal, the charge current 
would flow only in one spin channel and no charge current in the other spin channel, thus resulting in $\sigma^{A}_{xy}$ 
being twice as large as $\sigma^{S}_{xy}$. Therefore, Co$_{2}$FeIn and Co$_{2}$FeAl$_{0.5}$Si$_{0.5}$ may be
called anomalous Hall half-metals\cite{Tung13}, even though their electronic states near $E_F$ are far from fully spin-polarized
(see $P^D$ in Table II). Finally, we note that the ratio of spin current to charge current 
for the Hall current ($\eta = |P^H|$) in Co$_{2}$FeAl is large with $\eta > 150$ \%.

\subsection{Current spin polarizations}
The calculated spin polarizations of Hall ($P^H$) and longitudinal ($P^L$) currents as well as electronic states
at $E_F$ ($P^D$) for all the Heusler componds considered here are listed in Table II. Also listed in Table II are
the spin-decomposed Hall conductivities ($\sigma^{H\uparrow}_{xy}$ and $\sigma^{H\downarrow}_{xy}$) 
obtained using Eqs. (4) and (5). Remarkably, Table II shows that the calculated $P^L$ is nearly 100 \% in 
Co$_2$FeAl, Co$_2$FeGa and Co$_2$FeAl$_{0.5}$Si$_{0.5}$ even though their $P^D$ is significantly smaller than 100 \%.
This finding, therefore, indicates that these Heusler compounds are half-metallic from the viewpoint 
of charge transport, even though their electronic band structures are not. 
All the other compounds also have a high current spin polarization with $P^L > 85$ \%.
Therefore, all the Heusler compounds considered here may find valuable applications in spintronic devices.
Interestingly, Table II also demonstrates that
the $P^L$ and $P^D$ in Co$_2$FeSi, Co$_2$FeGe and Co$_2$FeSn could even have opposite signs. 
The calculated current spin polarization $P^L$ in Co$_2$FeSi and Co$_2$FeGe is positive, 
being in good agreement with recent spin Hall effect experiments\cite{Oki12}.
In contrast, the static spin polarization ($P^D$) 
differs from the experimental spin polarization even in sign (Table II). 
This clearly urges one to compare the measured 
spin polarization from transport experiments to the theoretical current spin polarization rather than
the static spin polarization which has often been done in the past.
  
The interesting finding that the $P^L$ and $P^D$ in Co$_2$FeSi, Co$_2$FeGe and Co$_2$FeSn differ in sign,
could be explained in terms of the calculated band structures. Figure 2(c) indicates that in Co$_2$FeSi,
for the spin-up channel, the $E_F$ cuts through the highly dispersive Co/Fe $spd$ and Si $sp$ hybridized bands.
On the other hand, for the spin-down channel, the $E_F$ is located at the bottom of the Co/Fe $d (e_g)$ dominated bands.
Consequently, the spin-down DOS at $E_F$ is higher than the spin-up DOS (see Fig. 1c and Table I), giving rise
to the negative value of $P^D$. From transport viewpoint, however, the spin-down Co/Fe $d (e_g)$ dominated bands 
which are narrow (Fig. 2c), would have large effective masses and small Fermi velocities, 
thereby contributing little to the charge current.
On the other hand, the spin-up Co/Fe $spd$ and Si $sp$ hybridized bands which are highly dispersive, would have
large Fermi velocities and small effective masses, thus providing the dominant contribution to the charge current.
Therefore, the current spin polarization is positive, being in good agreement with the experiments (Table II).   

\begin{figure}
\includegraphics[width=8cm]{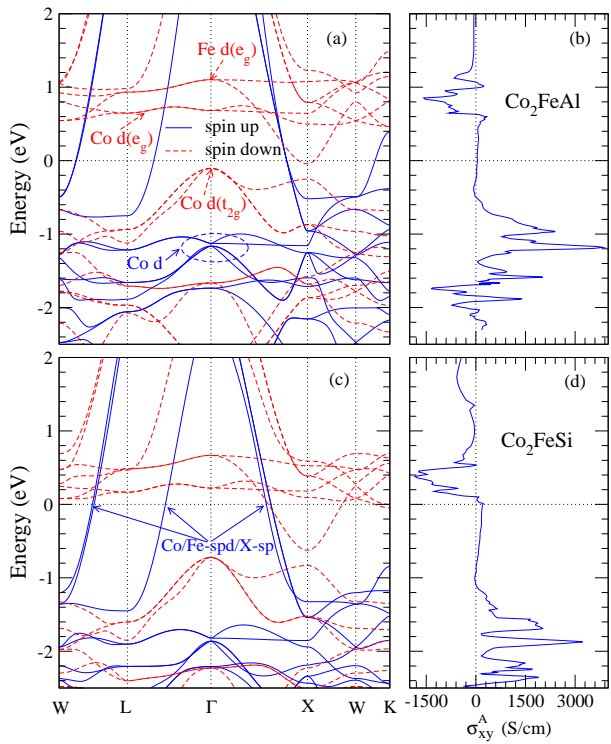}\
\caption{(color online) Scalar relativistic band structure [(a) and (c)], 
and anomalous Hall conductivity ($\sigma^A_{xy}$) 
[(b) and (d)] for Co$_2$FeAl, and Co$_2$FeSi, respectively. The Fermi energy is at zero.}
\label{sigma1}
\end{figure}

Many experiments\cite{Makinistian13,Karthik07,Karthik07-1,Gercsi06,Nakatani07,USPAT13}
especially PCAR measurements on Co$_2$Fe-based Heusler alloys
have been carried out to determine their spin polarization which is a key factor
for their spintronic applications. Majority of these experiments were focused on Co$_2$FeSi 
mainly because highly L2$_1$ ordered Co$_2$FeSi samples could be fabricated.
However, the $P^{L}$ values derived from PCAR experiments on Co$_2$FeSi vary significantly
from 45 \% to 60\% (Table II), depending on the quality of the samples. This could be expected
because the spin polarization determined by a PCAR experiment depends not only on
the degree of the ordering and the defects in the sample but also on the quality
of the contact and the substrate\cite{Hsu12}. Nevertheless, the theoretical $P^L$ value of 86 \%
agrees rather well with the experimental value of 80\% from the  
nonlocal spin-valve experiment\cite{Hamaya12} on highly L2$_1$-ordered specimens.
However, the measured $P^{L}$ values for Co$_2$FeAl, Co$_2$FeGa and Co$_2$FeAl$_{0.5}$Si$_{0.5}$ 
are around 60 \%, which is far from the predicted $P^{L}$ value of $\sim$100 \% for these compounds (Table II).
These significant discrepancies may reflect the fact that the samples used in the experiments\cite{Karthik07-1,USPAT13,Karthik07}
had a high degree of the B2 antisite disorders.

\subsection{Effects of on-site Coulomb interaction}

\begin{table*}
\caption{Total magnetic moment ($m^{tot}$), total spin magnetic moment ($m_s^{tot}$) ($\mu_B$/f.u.), 
spin-decomposed density of states at the Fermi level
[$N^{\uparrow}(E_F)$, $N^{\downarrow}(E_F)$] (states/eV/f.u.),  spin polarization of the electronic states at the Fermi level $P^{D}$ ($\%$), longitudinal current polarization $P^L$ ($\%$, anomalous [$\sigma_{xy}^A$ (S/cm)] 
and spin [$\sigma_{xy}^S$ ($\hbar$S/e cm)] Hall conductivities and Hall current spin polarization $P^H$ ($\%$)
of the Co$_2$FeX Heusler compounds from both the GGA and GGA+$U$ calculations. 
The on-site Coulomb (exchange) interaction $U$ ($J$) for Co and Fe used are 2.82 (0.9) eV and 2.6 (0.8) eV, respectively.
}
\begin{center}
\begin{tabular}{ccc@{\hspace{0.2cm}}c@{\hspace{0.5cm}}c@{\hspace{0.4cm}}c@{\hspace
{0.5cm}}c@{\hspace{0.5cm}}c@{\hspace{0.5cm}}c@{\hspace{0.5cm}}c@{\hspace{0.5cm}}c}\hline\hline
Co$_2$FeX & &$m^{tot}$ &$m_s^{tot}$ & $N^{\uparrow}(E_F)$ & $N^{\downarrow}(E_F)$ & $P^{D}$ & $P^L$ & $\sigma^{A}_{xy}$ &$\sigma^{S}_{xy}$
 &$P^H$ \\ \hline
Co$_2$FeAl &GGA     & 5.123 & 4.993 & 0.862 & 0.059 & 87 & 100  & 39 & 35 & -180 \\
           &GGA+$U$ & 5.202 & 4.999 & 0.753 & 0.003 & 99 & 100  & 98 & 69 & -140 \\
Co$_2$FeGa &GGA     & 5.149 & 5.016 & 0.885 & 0.189 & 65 & 98   &181 & 56 & -62  \\
           &GGA+$U$ & 5.259 & 5.043 & 0.772 & 0.515 & 20 & 100  & 89 & 67 & -151 \\
Co$_2$FeAl$_{0.5}$Si$_{0.5}$&GGA     & 5.523 & 5.376 & 0.755 & 0.399 & 31 &  92 & 124 & 74 & -119 \\
                            &GGA+$U$ & 5.700 & 5.498 & 0.667 & 0.001 & 100& 100 & 139& 87 & -125 \\
Co$_2$FeSi &GGA     & 5.688 & 5.541 & 0.714 & 2.476 & -55 &  86 &189 & 24 & -25 \\
           &GGA+$U$ & 6.196 & 5.998 & 0.587 & 0.008 & 98 & 100  & 73 & 54 & -148 \\
Co$_2$FeGe &GGA     & 5.854 & 5.693 & 0.785 & 2.288 &-49 &  89  &119 &-29 & -49 \\
           &GGA+$U$ & 6.222 & 5.997 & 0.624 & 0.003 & 99 & 100  & 14 & 40 & -570 \\
\hline\hline

\end{tabular}
\end{center}
\end{table*}

To examine the effect of on-site Coulomb interaction,
we further perform the calculations in the GGA+$U$ scheme\cite{Lie95}. 
The on-site Coulomb repulsion $U$(exchange interaction $J$) used are 2.82 (0.9) 
and 2.6 (0.8) eV for Co and Fe, respectively, which are widely used 
for Co$_2$-based Heusler compounds.\cite{Gerhard07}
The results from these GGA+$U$ calculations are compared with those of the GGA calculations 
in Table III. We notice that the total spin magnetic moments ($m_s^{tot}$) from the GGA and GGA+$U$ calculations  
are almost identical in all the Heusler compounds except Co$_2$FeSi and Co$_2$FeGe. This may be expected
since the GGA $m_s^{tot}$ is already nearly saturated in these compounds. Including the on-site Coulomb interaction
increases the $m_s^{tot}$ in Co$_2$FeSi and Co$_2$FeGe to the saturation values. 
Note that the measured magnetic moments should be compared with the calculated total magnetic moments ($m^{tot}$)
instead of total spin magnetic moments ($m_s^{tot}$) in Tables I and III.
The $m^{tot}$ contains both the $m_s^{tot}$ and the total orbital magnetic moment which cannot be neglected
in the Heusler compounds studied here because the orbital magnetic moments on the Fe and Co atoms are 
rather significant (Table I). Tables I and III together show that including the on-site Coulomb interaction
actually increases the small discrepancies between the experiments and the GGA calculations
found in all the Heusler compounds except Co$_2$FeSi where the difference of 0.3 $\mu_B$/f.u. is reduced slightly to 0.2 $\mu_B$
per formula unit.

Table III shows that the on-site Coulomb interaction has a pronounced effect on
the spin polarization. First, the current spin polarization ($P^L$) for all the studied compounds
is now 100 \% from the GGA+$U$ calculations. Second, the static spin polarization 
($P^D$) approaches to 100\% for all the compounds except Co$_2$FeGa.
Therefore, these Heusler compounds become half-metals in terms of both the band structure
and current spin polarization. This may be expected because the main effect of the on-site Coulomb interaction 
is to raise the spin-down Fe and Co $d$-dominant conduction bands. Consequently, if sufficiently large $U$
values are used, the spin-down Fe and Co $d$-dominant conduction bands will move to above $E_F$.
This will open a gap in the spin-down channel and thus give rise to zero spin-down DOS at $E_F$ (Table III). 
However, the spin-down GGA+$U$ band gap is as large as 0.9 eV in Co$_2$FeSi, for example,
being nearly 100 times larger than the measured one\cite{Dirk13}.  
Interestingly, including Coulomb $U$ changes the spin polarization $P^D$ 
in Co$_2$FeSi and Co$_2$FeGe from negative to positive (Table III). 
However, it should be emphasized that the mechanism of the spin polarization sign change 
here is very different from the sign difference between the $P^D$ and $P^L$
in the GGA calculations. Nevertheless, whether the $P^D$ is positive or negative can be
tested by spin-polarized angle-resolved photoemission experiments which, unfortunately,
have not been reported on any Heusler compound studied here.  

Table III also indicates that including the on-site Coulomb $U$ changes the calculated 
AHC and SHC substantially. In particular, the $\sigma^{A}_{xy}$ 
gets reduced significantly for all the studied compounds except Co$_2$FeAl (Table III).
For example, the theoretical $\sigma^{A}_{xy}$ for Co$_2$FeGe is 119 S/cm from the GGA calculation
but is reduced to 14 S/cm when the on-site Coulomb $U$ is included.
This suggests that by comparing the calculated $\sigma^{A}_{xy}$ with the measured one,
one could assess whether or not including on-site Coulomb $U$ is needed to properly describe 
the electronic properties of a Co$_2$Fe-based Heusler compound. 
The measured $\sigma^{A}_{xy}$ of Co$_2$FeSi\cite{Dirk13} is $\sim$160 S/cm, being in
good agreement with the GGA result (Table II). However, it is two times larger than
the result of the GGA+$U$ calculation (about 70 S/cm). This indicates that Co$_2$Fe-based
Heusler compounds are not strongly correlated systems and there may be no
need to include the on-site Coulomb $U$ for these compounds. 

\section{Conclusions}
We have carried out a systematic {\it ab initio} study of the anomalous Hall effect 
and current spin polarization as well as the magnetic properties of the Co$_2$FeX (X = Al, Ga, In, Si, Ge, Sn) 
Heusler compounds in the cubic L2$_1$ structure by using 
the highly accurate all-electron FLAPW method.
First, we find that the spin-polarization of the longitudinal current ($P^L$) in Co$_2$FeX (X = Al, Ga
and Al$_{0.5}$Si$_{0.5}$) is $\sim$100 \% even though the static spin polarization ($P^D$) is not. 
Furthermore, the other compounds also have a high current
spin polarization with $P^L > 85$ \%. This indicates that all the Co$_2$FeX compounds 
are promising for spintronic devices. Interestingly, $P^D$ is negative in Co$_2$FeX (X = Si, Ge and Sn),
differing in sign from the $P^L$ as well as from that from the transport experiments. Second, the calculated 
AHCs are moderate, being within 200 S/cm, and agree well with the available experiments
on highly L2$_1$ ordered Co$_2$FeSi specimen although they differ significantly from the reported experiments
on other compounds where the B2 antisite disorders were present. Surprisingly, the AHC in Co$_2$FeSi
decreases and then changes from the negative to positive when Si is replaced by Ge and finally by Sn.
Third, the calculated total magnetic moments are in good agreement with the experiments in all the
studied compounds except Co$_2$FeSi where a difference of 0.3 $\mu_B$/f.u. exists.
We have also performed the GGA+$U$ calculations in order to examine the effects of the on-site Couloumb repulsion.
We find that including the $U$ changes 
the calculated total magnetic moment, spin polarization and AHC significantly. In most cases, unfortunately,
this results in a worse agreement with the available experimental results. These interesting findings
are analyzed in terms of the underlying band structures.

\section*{Acknowledgments}
The authors acknowledge supports from the Ministry of Science and Technology and also 
the Academia Sinica of the ROC.
They also thank the NCHC of Taiwan for providing CPU time.


\begin{references}
\bibitem{note2} The experimental $b$ values\cite{Dirk13,Imort12,Vil11,Wu13} listed in Table II were obtained by fitting
the expression $\rho_A =a\rho_{xx}+b\rho_{xx}^2$ to the measured anomalous Hall resistivity $\rho_A$.
The coefficient $b$ is the scattering-independent term containing both the intrinsic and extrinsic side-jump
contributions while the coefficient $a$ is the extrinsic skew-scattering term\cite{Nagaosa10}.
\end{references}
\end{document}